\documentclass[12pt,a4paper,final]{iopart}
\usepackage{iopams}

\usepackage{graphicx}
\usepackage{dcolumn}
\usepackage{float}
\usepackage{iopams}
\usepackage{multirow}
\usepackage[breaklinks=true,colorlinks=true,linkcolor=blue,urlcolor=blue,citecolor=blue]{hyperref}

\begin{document}

\title{Reverse process of usual optical analysis of boson-exchange superconductors: impurity effects on $s$- and $d$-wave superconductors}

\author{Jungseek Hwang}

\address{Department of Physics, Sungkyunkwan University, Suwon, Gyeonggi-do 440-746, Republic of Korea}

\ead{jungseek@skku.edu}
\date{\today}

%\maketitle

%
% Abstract
%
\begin{abstract}
We performed a reverse process of a usual optical data analysis of boson-exchange superconductors. We calculated the optical self-energy from two (MMP and MMP+peak) input model electron-boson spectral density functions using Allen's formula for one normal and two ($s$- and $d$-wave) superconducting cases. We obtained the optical constants including the optical conductivity and the dynamic dielectric function from the optical self-energy using an extended Drude model, and finally calculated reflectance spectrum. Furthermore to investigate impurity effects on optical quantities we added various levels of impurities (from the clean to the dirty limit) in the optical self-energy and performed the same reverse process to obtain the optical conductivity, the dielectric function, and reflectance. From these optical constants obtained from the reverse process we extracted the impurity dependent superfluid densities for two superconducting cases using two independent methods (the Ferrel-Glover-Tinkham sum rule and the extrapolation to zero frequency of $-\epsilon_1(\omega)\omega^2$); we found that a certain level of impurities is necessary to get a good agreement on results obtained by the two methods. We observed that impurities give similar effects on various optical constants of $s$- and $d$-wave superconductors; the more impurities give the more distinct gap feature and the less superfluid density. However, the $s$-wave superconductor gives the superconducting gap feature more clearly than the $d$-wave superconductor because in the $d$-wave superconductors the optical quantities are averaged over the anisotropic Fermi surface. Our results also supply helpful information to see how characteristic features of the electron-boson spectral function and the $s$- and $d$-wave superconducting gaps appear in various optical constants including raw reflectance spectrum. Our study also may help to understand the usual optical analysis process thoroughly. Further systematic study of experimental data collected at various conditions using the optical analysis process will help to reveal the origin of the mediated boson in the boson-exchange superconductors.
\end{abstract}

\pacs{74.25.Gz, 74.25.F, 74.25.-q}

\maketitle

%
%\bigskip

\section{Introduction}

Strongly correlated electron systems including high temperature superconductors have been very intriguing research objects and been studied intensively by various experimental techniques and theoretical methods\cite{imada:1998,carbotte:2011}. Optical and/or infrared spectroscopic techniques have been playing an important role to delve into the origin of the correlations among electrons in the correlated systems\cite{basov:2005}. In general measured reflectance spectra of correlated electron systems including boson-exchange superconductors\cite{carbotte:1990} can be analyzed, as follows: First a Kramers-Kronig analysis\cite{wooten} is performed to obtain the optical constants including the optical conductivity. Then an extended Drude model is applied to get the optical self-energy\cite{hwang:2004} which can carry information of the correlations and is in one-to-one correspondence with the well-known quasiparticle self-energy\cite{damascelli:2003,carbotte:2005,hwang:2007b}. An extended Allen's formulas\cite{allen:1971,mitrovic:1985,shulga:1991,sharapov:2005} are used for getting more fundamental quantity, the electron-boson spectral density function, from the optical self-energy. The Allen's formulas relate linearly the optical self-energy to the electron-boson spectral density function. The electron-boson spectral density function can be obtained usually by solving numerically\cite{dordevic:2005,schachinger:2006,heumen:2009} the extended Allen's formula which is an integral equation. Further systematic optical studies provide temperature and doping dependent properties of the electron-boson spectral density function of high temperatures superconductors\cite{hwang:2006,hwang:2007}.

It has been known that the electron-boson spectral functions of cuprates obtained from various spectroscopic experimental [angle resolved photoemission (ARPES)\cite{valla:2007}, (scanning) tunneling (STS)\cite{zasadzinski:2006,ahmadi:2011}, inelastic neutron scattering (INS), Raman\cite{muschuler:2010}, and infrared (IR)\cite{hwang:2006,hwang:2007,hwang:2008c}] techniques consist of two components: one broad background and a sharp mode. While the background shows small doping dependence the sharp mode shows strong doping and temperature dependencies\cite{carbotte:2011}. The sharp mode appears and grows with decreasing temperature by consuming the spectral weight of the high temperature broad background\cite{carbotte:2011,hwang:2004,hwang:2006,hwang:2007,zasadzinski:2006,ahmadi:2011,muschuler:2010,dai:1999,johnson:2001,sato:2003,hwang:2007a,zhang:2008,yang:2009}. The sharp mode is known to be associated with the well-known magnetic resonance mode\cite{rossat:1991}. As reducing the sample temperature the sharp mode starts to appear above the superconducting transition temperature ($T_c$) in underdoped cuprates and at $T_c$ in the optimally and overdoped cuprates\cite{carbotte:2011,hwang:2004,dai:1999,johnson:2001,sato:2003}.
Many researchers believe that the high temperature superconductors including cuprates are boson-exchange superconductors as the conventional phonon-mediated superconductors and the electron-boson spectral functions can carry information of the bonding force of Cooper pairs in the high temperature superconductors. So the electron-boson spectral density function observed in cuprates has been known as the superconducting pairing glue function\cite{hwang:2004,zasadzinski:2006,hwang:2008c,carbotte:1999,heumen:2009,maier:2008,dahm:2009,kyung:2009}. Figuring out the microscopic origin of the boson which involves in the bonding force is one of the most important problems in contemporary condensed matter physics. We believe that the electron-boson density functions obtained by various experimental techniques will play a very important role to solve the problem.

In this paper we performed a reverse process of the usual optical analysis process described above to understand better the usual process and investigate impurity effects on optical quantities. We started from an impurity-free pure (or ideal) system, and then introduced elastic impurities into the system, and finally examined the impurity effects on the optical constants including reflectance. To achieve this result we introduced two input model electron-boson spectral density functions, which are modeled with general shapes of the electron-boson spectral density functions observed by various spectroscopic experiments\cite{carbotte:2011}, and then calculated the imaginary part of the optical self-energy (or the optical scattering rate) from the input spectral density functions using the Allen's formulas\cite{allen:1971,schachinger:2006}. One of the two input model spectral density functions is the phenomenological antiferromagnetic spin fluctuation\cite{millis:1990}, which is proposed by Millis, Monien, and Pines (MMP) and is also known as an MMP mode. The other is a combination of the MMP mode and a sharp Gaussian mode which is modeled of an observed electron-boson spectral function in cuprate system at low temperature below the superconducting transition temperature\cite{carbotte:2011,hwang:2006}. We also considered for three different cases: one normal and two different ($s$- and $d$- wave) types of superconducting states to investigate superconducting gap symmetry effects. Furthermore we used a Kramers-Kronig relation to obtain the real part of the optical self-energy from the imaginary part (or the optical scattering rate). We calculated the optical conductivity using the extended Drude formalism\cite{hwang:2004,puchkov:1996}. We also calculated the dynamic dielectric function using the relations between optical constants and then finally obtained the normal incident reflectance spectrum using the Fresnel equations.

Impurities are not avoidable experimentally, especially for doped materials with substitution. The impurity effect on physical quantities is an important physical issue especially for unconventional superconductors including $d$-wave superconductors\cite{liang:1994,sun:1995}. Most experimental results may be contaminated by the impurities in the material systems or sometimes impurities are introduced experimentally to study their effects on the physical quatities\cite{janossy:1994}. There may be physical issues, which ones may need to be aware of impurity effects, for analysis of optical spectra. For example violation of the Ferrel-Glover-Tinkham (FGT) sum rule\cite{glover:1956,ferrell:1958} in certain superconductors can be interpreted as the lowering kinetic energy which comes from contributions of both the condensation energy and the potential energy increase due to larger Coulomb repulsion in the Cooper pairs\cite{hirsch:1992,hirsch:2000,norman:2003}. So we added several different impurity scattering rates (from the clean to the dirty limit) to the impurity-free optical scattering rate. We performed the same reverse procedure to obtain the optical conductivities, the dynamic dielectric constants and reflectance spectra from the optical self-energies which contain impurities. We discuss impurity effects on other optical constants including reflectance. Our results show that measured reflectance spectrum and other optical constants clearly show important characteristic features of the input electron-boson spectral function and the superconducting gap feature. This may be quite useful because one can see important features even in raw reflectance spectra without going through a full optical analysis process. And then we investigate impurity effects on the physical quantities including the partial sum, spectral weight redistribution by the Holstein process, and the superfluid density which can be obtained from the optical conductivity and the dielectric function. Our results show that impurities may affect on the FGT sum rule; one may need enough level of impurities in the material systems to hold the FGT sum rule. We observed that the absolute superfluid density depends on the impurity level; the more impurities give the less superfluid density. We also compared our results obtained by the reverse process with experimental data of optimally doped Bi$_2$Sr$_2$CaCu$_2$O$_{8+\delta}$ and observed good agreements for all optical constants considered including reflectance. These good agreements confirm that our reverse process is reliable. We could understand thoroughly the usual optical process through this study. We believe that further study in this direction may help to reveal the microscopic origin of the exchange boson in the high temperature superconductors as in the conventional superconductors by collecting more experimental data and applying the optical analysis process to them to obtain various physical properties of the exchange boson.

This paper is structured with the following sections. In session 2 we introduce general theoretical formalisms which we used in this paper. In section 3 we show and discuss our results obtained using the formalisms. In section 4 we summarize our results with some useful comments.

\section{Theoretical formalisms}

Theoretical analysis methods of measured optical data of $s$-wave and $d$-wave superconductors have been developed considerably\cite{carbotte:2011,carbotte:1990}. One of important progresses, which has been successful for conventional superconductors\cite{farnworth:1974,farnworth:1976}, is the boson-exchange model approach to study unconventional superconductors\cite{carbotte:2011}. One can extract the electron-boson spectral density function from measured optical self-energy\cite{hwang:2006,hwang:2007,schachinger:2000,schachinger:2003}. The optical self-energy [$\tilde{\Sigma}^{op}(\omega)$] can be defined through an extended Drude model\cite{hwang:2004}. In the extended Drude model the scattering is inelastic ({\it i.e.} frequency dependent) while in the simple Drude model the scattering is elastic ({\it i.e.} frequency independent). When the scattering rate is frequency dependent its Kramers-Kronig counterpart is not zero any more. So we need a complex optical quantity to describe the inelastic scattering. The complex optical quantity is called the optical self-energy [$\tilde{\Sigma}^{op}(\omega) = \Sigma^{op}_1(\omega) + i\Sigma^{op}_2(\omega)$]. The optical self-energy is a corresponding optical quantity of the quasiparticle self-energy [$\tilde{\Sigma}^{qp}(\omega)$]\cite{hwang:2007b}. The imaginary part of the optical self-energy is related to the optical scattering rate [$1/\tau^{op}(\omega)$ or the frequency dependent scattering rate]. The real part is related to the mass enhancement factor [$\lambda^{op}(\omega)$] which is caused by the correlation among electrons\cite{hwang:2004}.

Here we introduce formalisms which we used in this paper. The formalisms are a series of equations of the reverse process of the usual optical data analysis. We start from a model electron-boson spectral density function [$I^2\chi(\omega)$] and use the Allen's fromula\cite{allen:1971,schachinger:2006} to obtain the imaginary part of the optical self-energy [$\Sigma^{op}_2(\omega)$]. The Allen's formula is an integral equation which relates linearly the electron-boson spectral density function [$I^2\chi(\omega)$] to the imaginary part of the optical self-energy [$\Sigma^{op}_2(\omega)$]. The Allen's formula can be described as follows:
\begin{equation}\label{eq1}
-2\Sigma^{op}_2(\omega) \equiv \frac{1}{\tau^{op}(\omega)} = \int_0^{\infty}d\Omega \: I^2\chi(\Omega) \:K(\omega,\Omega)+\frac{1}{\tau^{op}_{imp}(\omega)}
\end{equation}
where $1/\tau^{op}(\omega)$ is the optical scattering rate, $K(\omega,\Omega)$ is the kernel for the Allen's integral equation and $1/\tau^{op}_{imp}(\omega)$ is the impurity scattering rate. We have different kernels for the normal state (here $T =$ 0 case) and $s$- and $d$-wave superconducting states. The kernels can be written as follows:
\begin{eqnarray}\label{eq2}
K(\omega,\Omega) &=& 2\pi\Big{(}1-\frac{\Omega}{\omega}\Big{)}\Theta(\omega-\Omega) \:\:\:(\mbox{for normal state}) \nonumber \\ &=& 2\pi\Big{(}1-\frac{\Omega}{\omega}\Big{)}\Theta(\omega-2\Delta_0 - \Omega) \nonumber \\ &&\times E\Big{(}\sqrt{1-\frac{4\Delta_0^2}{(\omega-\Omega)^2}}\Big{)} \:\:\:(\mbox{for $s$-wave SC }) \nonumber \\
&=& 2\pi\Big{(}1-\frac{\Omega}{\omega}\Big{)}\Big{\langle}\Theta(\omega - 2\Delta_0(\theta) - \Omega) \nonumber \\ &&\times E\Big{(}\sqrt{1\!-\!\frac{4\Delta_0(\theta)^2}{(\omega-\Omega)^2}}\Big{)} \! \Big{\rangle}_\theta (\mbox{for $d$-wave SC})
\end{eqnarray}
where $\Theta(x)$ is the Heaviside step function ({\it i.e.} 1 for $x \geq 0$ and 0 for $x < 0$), $E(x)$ is the complete elliptic integral of the second kind, and $\langle \cdots \rangle_\theta$ stands for the angular average over $[0, \pi/4]$. $\Delta_0$ and $\Delta_0(\theta) [= \Delta_0 \cos(2\theta)]$ are the superconducting gaps for $s$- and $d$- wave symmetry cases, respectively.

The impurity scattering rates [$1/\tau^{op}_{imp}(\omega)$] for normal and $s$- and $d$- wave superconducting states can be described as follows\cite{allen:1971,schachinger:2006}:
\begin{eqnarray}\label{eq3}
\frac{1}{\tau^{op}_{imp}(\omega)} &=& \frac{1}{\tau_{imp}} \:\:\:(\mbox{for normal state}) \nonumber \\ &=& \frac{1}{\tau_{imp}} E\Big{(}\sqrt{1-\frac{4\Delta_0^2}{\omega^2}}\Big{)} \:\:\:(\mbox{for $s$-wave SC }) \nonumber \\
&=&  \!\!\frac{1}{\tau_{imp}}\Big{\langle}\! E\Big{(}\sqrt{1\!-\!\frac{4\Delta_0(\theta)^2}{\omega^2}}\Big{)} \! \Big{\rangle}_\theta (\mbox{for $d$-wave SC})
\end{eqnarray}
where $1/\tau_{imp}$ is the impurity scattering rate, which is a constant. We note that for the superconducting cases the impurity scattering rates are frequency dependent.

In the extended Drude model formalism\cite{hwang:2004,puchkov:1996}, which can describe correlated charge carriers, we can relate the optical conductivity, $\tilde{\sigma}(\omega)[=\sigma_1(\omega) + i\sigma_2(\omega)$] to the optical self-energy, $\tilde{\Sigma}^{op}(\omega) [= \Sigma^{op}_1(\omega) + i\Sigma^{op}_2(\omega)]$ as follows\cite{hwang:2004}:
\begin{eqnarray}\label{eq4}
\tilde{\sigma}(\omega) &=& i\frac{\Omega_p^2}{4 \pi}\:\frac{1}{\omega -2\tilde{\Sigma}^{op}(\omega)} \:\:\:\: \mbox{or} \nonumber \\ -2\tilde{\Sigma}^{op}(\omega) &=& i\frac{\Omega_p^2}{4 \pi}\:\frac{1}{\tilde{\sigma}(\omega)}-\omega
\end{eqnarray}
where $\Omega_p$ is the plasma frequency of charge carriers. The imaginary and real parts of the optical self-energy can be related to respectively the optical scattering rate [$1/\tau^{op}(\omega)$] and the optical mass enhancement factor [$\lambda^{op}(\omega)$] as $-2\Sigma^{op}_2(\omega) = 1/\tau^{op}(\omega)$ and $-2\Sigma^{op}_1(\omega) = \lambda^{op}(\omega)\omega \equiv [m^*(\omega)/m_e -1]\omega$, where $m^*(\omega)$ and $m_e$ are the enhanced mass from the correlation and the electron mass, respectively. We note that the real and imaginary parts of the optical self-energy form a Kramers-Kronig pair. Therefore if you know one of them we can get the other using the Kramers-Kronig relation between them.

In normal state at $T =$ 0, $-2\Sigma^{op}_1(\omega)$ can be written analytically as follows\cite{carbotte:2005,allen:1971,hwang:2008b}:
\begin{equation}\label{eq4}
-2\Sigma^{op}_1(\omega) = 2\int_{0}^{\infty}d\Omega I^2\chi(\Omega) \Big{[} \frac{\Omega}{\omega}\ln{\Big{|} \frac{\Omega^2-\omega^2}{\Omega^2}\Big{|}} + \ln{\Big{|} \frac{\Omega+\omega}{\Omega-\omega}\Big{|}}\Big{]}.
\end{equation}
We also introduce an interesting relation at zero frequency as follows:
\begin{eqnarray}\label{eq6}
\lambda^{op}(0) &=&  \lim_{\omega \rightarrow 0} \frac{-2\Sigma^{op}_1(\omega)}{\omega} = \lim_{\omega \rightarrow 0} \frac{d[-2\Sigma^{op}_1(\omega)]}{d\omega} \:\: \nonumber \\
&=& 2\int_0^{\omega_{c}} d\Omega \frac{I^2\chi(\Omega)}{\Omega} \equiv \lambda.
\end{eqnarray}
The second step in the first equation is obtained by applying the L'Hopitals's rule. The last equation is called the coupling constant which we denote as $\lambda$ (see in Fig. \ref{fig5}) and $\omega_{c}$ is the cutoff frequency.

The dynamic dielectric function [$\tilde{\epsilon}(\omega)= \epsilon_1(\omega)+i\epsilon_2(\omega)$] can be related to the optical conductivity as follows:
\begin{equation}\label{eq7}
\tilde{\epsilon}(\omega) = \epsilon_{H} + i\frac{4 \pi}{\omega}\tilde{\sigma}(\omega)
\end{equation}
where $\epsilon_{H}$ is the high frequency background dielectric constant.

Finally one can calculate the single bounce reflectance [$R(\omega)$] at normal incidence from the dynamic dielectric function using the Fresnel equations and the relations between optical constants\cite{wooten}. The reflectance can be described as follows:
\begin{eqnarray}\label{eq8}
R(\omega) &=& \Big{|} \frac{\tilde{N}(\omega)-1}{\tilde{N}(\omega)+1} \Big{|}^2 = \Big{|} \frac{\sqrt{\tilde{\epsilon}(\omega)}-1}{\sqrt{\tilde{\epsilon}(\omega)}+1} \Big{|}^2 \:\: \nonumber \\ &=& \frac{\sqrt{\epsilon_1^2 + \epsilon_2^2} - \sqrt{2 (\epsilon_1 + \sqrt{\epsilon_1^2 + \epsilon_2^2})} + 1}
{\sqrt{\epsilon_1^2 + \epsilon_2^2} + \sqrt{2 (\epsilon_1 + \sqrt{\epsilon_1^2 + \epsilon_2^2})} + 1}.
\end{eqnarray}
where $\tilde{N}$ is the complex index of refraction. $\tilde{N} = n+i\kappa = \sqrt{\tilde{\epsilon}}$, where $n$ and $\kappa$ are called the index of refraction and the extinction coefficient, respectively. We note that if a sample is not transparent its reflectance is the single bounce one. Here 1 is the dielectric constant of the vacuum or air. In the following section we will show and discuss about the results of our calculations using the formulas of the reverse process.

\section{Results and discussions}

\subsection{Model electron-boson spectral density, $I^2\chi(\omega)$}

\begin{figure}[t]
  \vspace*{-0.3 cm}%
  \centerline{\includegraphics[width=4.2 in]{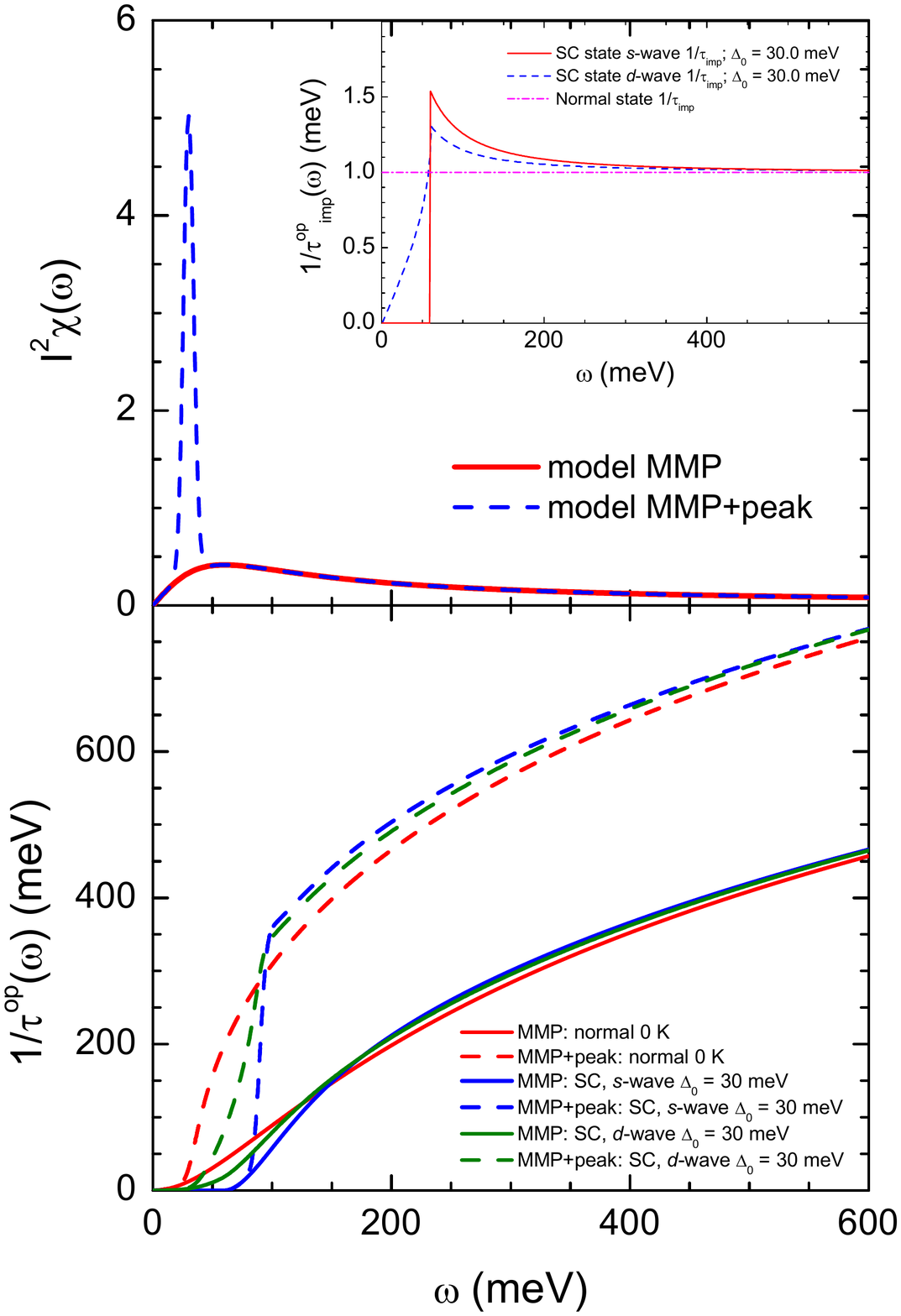}}%
  \vspace*{-0.7cm}%
\caption{(Color online) Two input model electron-boson spectral density functions (MMP and MMP+peak) in the upper frame, the calculated optical scattering rates for the three cases (one normal and two SC cases) in the lower frame, and impurity scattering rates for the three cases in the inset (see in the text).}
 \label{fig1}
\end{figure}

For our calculations of the reverse process we started with two input model electron-boson spectral density functions, $I^2\chi(\omega)$, which are based on experimental results of hole doped cuprates\cite{carbotte:2011}. One model function consists of a broad mode, of which shape is a phenomenological antiferromagnetic spin fluctuation proposed by Millis, Monien, and Pines\cite{millis:1990}. This is called the MMP model and described as $I^2\chi(\omega) = \frac{A_s \omega}{\omega^2 + \omega_{sf}^2}$ where $A_s$ (= 50 meV) and $\omega_{sf}$ (= 60 meV) are the amplitude and the characteristic frequency of the spin fluctuation mode, respectively. The MMP model has a maximum at $\omega_{sf}$ and is displayed in the upper frame (red solid line) of Fig. \ref{fig1}. The other model spectral density function consists of two components: the MMP and a relatively sharp Gaussian peak. The Gaussian peak is modeled as the magnetic resonance mode observed by an inelastic neutron scattering\cite{dai:1999,rossat:1991} and is also called as the optical resonance mode\cite{hwang:2004,hwang:2007}. We denote this model as an MMP+peak model and describe as $I^2\chi(\omega) = \frac{A_s \omega}{\omega^2 + \omega_{sf}^2} + \frac{A_p}{\sqrt{2 \pi} (d/2.35)}\exp{\Big{[}\!\!-\frac{(\omega-\omega_p)^2}{2 (d/2.35)^2}\Big{]}}$, where $A_p$ (= 50 meV), $\omega_p$ (= 30 meV), and $d$ (= 10 meV) are the amplitude, the peak frequency, and the width of the Gaussian mode, respectively. The MMP+peak model is also displayed in the upper frame (blue dashed line) of Fig. \ref{fig1} along with the MMP model (red solid line).

The physical quantities related to the electron-boson spectral density function are the coupling constant [$\lambda \equiv 2\int_0^{\omega_{c}} I^2\chi(\Omega)/\Omega \:d\Omega$, see also in Eq. (\ref{eq6})], the average electron-boson frequency ($\omega_{ln}$), and the maximum superconducting transition temperature ($T_c^{max}$). The maximum superconducting transition temperature can be calculated in a generalized McMillan formalism\cite{carbotte:1990,hwang:2008c}, which can be described as follows:
\begin{equation}\label{eq9}
k_B T^{max}_c \cong 1.13\: \hbar\: \omega_{ln}\: \exp{\Big{[}-\frac{1+\lambda}{\lambda}\Big{]}}
\end{equation}
where $k_B$ is the Boltzmann constant, $\hbar$ is the reduced Planck's constant, and $\omega_{ln}$ is the logarithmically averaged electron-boson frequency which is defined as $\omega_{ln} \equiv \exp{ [(2/\lambda)\int_0^{\omega_{c}} \ln{\Omega} \: I^2\chi(\Omega)/\Omega\:d\Omega ] }$ where $\omega_{c}$ is the cutoff frequency, we use $\omega_{c} =$ 600 meV. In Table \ref{table1} we show the calculated values of the coupling constant, the average electron-boson frequency, and the maximum superconducting transition temperature. We note that the superconducting transition temperature is influenced mostly by the average frequency. We also note that these values in the table are comparable to reported ones for cuprates\cite{hwang:2006,hwang:2007}.

\begin{table} [h]
 \begin{center}
 \resizebox{12 cm}{!} {
  \begin{tabular}{|c||c|c|c|}
    \hline
     Physical quantities & $  \lambda$ & $\:\:\:\:\:$ $\omega_{ln}$ (meV) $\:\:\:\:\:$ &$\:\:\:\:\:$ $T_c^{max}$ (K) $\:\:\:\:\:$ \\ \hline \hline
    MMP model & $\:\:\:\:\:$ 2.424 $\:\:\:\:\:$ & 50.67 & 161.8 \\ \hline
    MMP+peak model &$\:\:\:\:\:$ 5.830 $\:\:\:\:\:$& 36.61 & 148.8 \\
    \hline
  \end{tabular}
  }
 \end{center}
  \caption{The calculated coupling constant ($\lambda$), the average frequency ($\omega_{ln}$), and the maximum superconducting temperature ($T_c^{max}$) of our two model input electron-boson spectral density function (see in the text).}
  \label{table1}
\end{table}

\subsection{Reverse process}

Now we performed the reverse process starting from the two input electron-boson spectral density functions, $I^2\chi(\omega)$. We obtained the imaginary part of the optical self-energy (or the optical scattering rate) for three cases using Eq. (\ref{eq1}) and (\ref{eq2}) and the input electron-boson spectral density functions. The three cases are a normal (N) state at $T =$ 0 K and two superconducting (SC) states with $s$- and $d$- wave gap symmetries. The gaps at $T =$ 0 K are $\Delta_0$ (= 30 meV) and $\Delta_0 \cos{(2\theta)}$ for $s$- and $d$- wave SC cases, respectively. The obtained optical scattering rates [$1/\tau^{op}(\omega)$] for the three cases using the two different $I^2\chi(\omega)$ are displayed in the lower frame of Fig. \ref{fig1}. While the MMP model gives gradual increases for all three cases the MMP+peak shows sharp rises with onset frequencies near the Gaussian peak energy ($\omega_p =$ 30 meV) for the normal and the peak energy plus two times of the SC gap ($\omega_p + 2\Delta_0 =$ 90 meV) for the $s$- wave SC. For the $d$-wave SC case the sharp rise starts from near $\omega_p$ (30 meV) and ends at $\omega_p + 2\Delta_0$ ({\it i.e.} 90 meV). We also calculated the impurity scattering rates [$1/\tau^{op}_{imp}(\omega)$] for the three cases using Eq. (\ref{eq3}) with $1/\tau_{imp} =$ 1.0 meV and $\Delta_0 =$ 30 meV and displayed them in the inset of Fig. \ref{fig1}. While the normal impurity scattering rate is independent of frequency both SC impurity scattering rates show strong frequency dependencies below and above $2\Delta_0$ and then become the normal impurity scattering rate value at high frequency region above 500 meV.

\begin{figure}[t]
  \vspace*{-0.3 cm}%
  \centerline{\includegraphics[width=6.0 in]{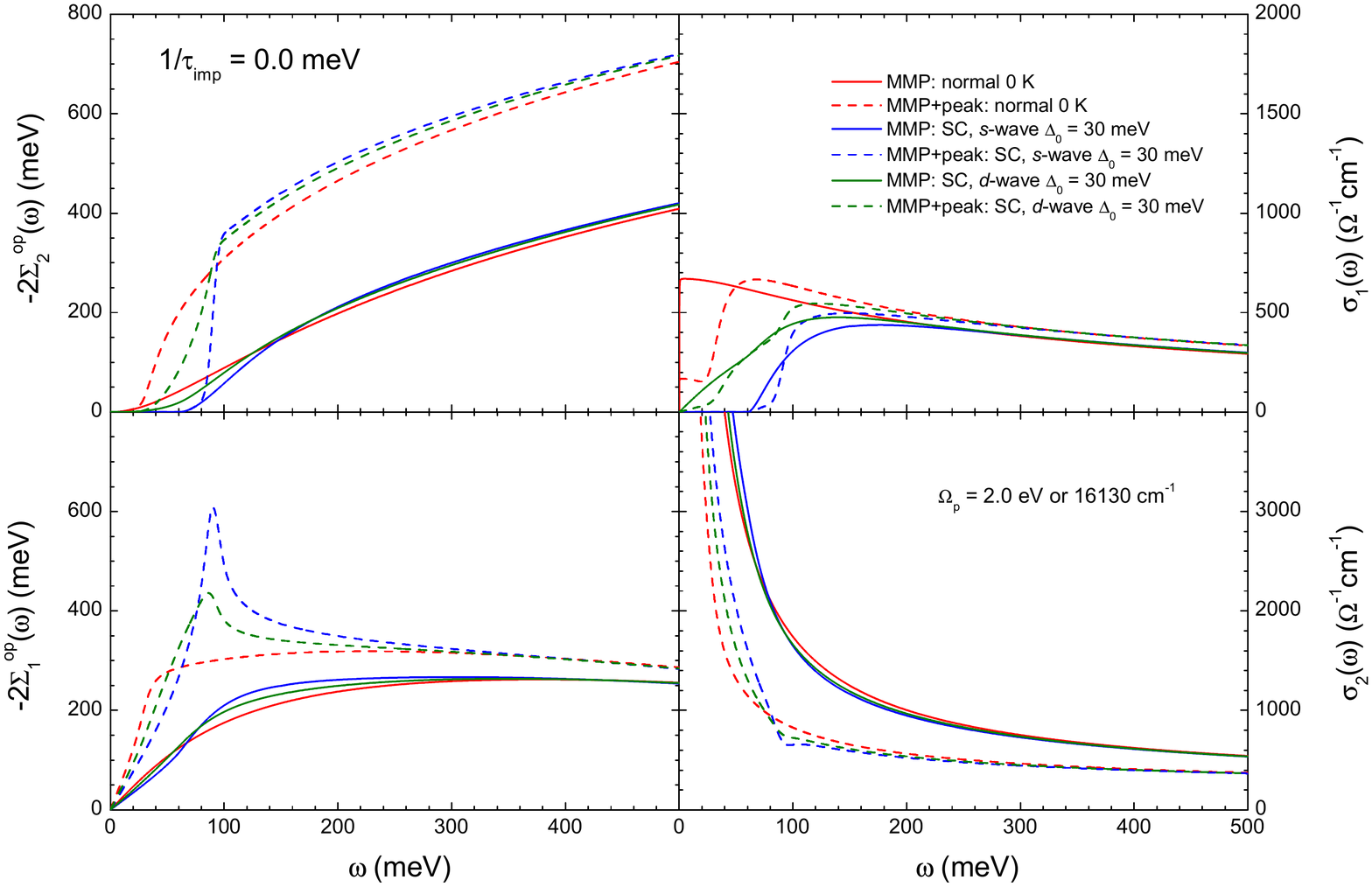}}%
  \vspace*{-0.7 cm}%
\caption{(Color online) The real (lower left frame) and imaginary (upper left frame) parts of the optical self-energy for all three cases are shown in the left frames. The real parts are obtained from corresponding imaginary parts using a Kramers-Kronig relation. The real (upper right frame) and imaginary (lower right frame) parts of the optical conductivity are obtained using the extended Drude formalism for all three cases and displayed in the right frames.}
 \label{fig2}
\end{figure}

\begin{figure}[t]
  \vspace*{-0.3 cm}%
  \centerline{\includegraphics[width=6.0 in]{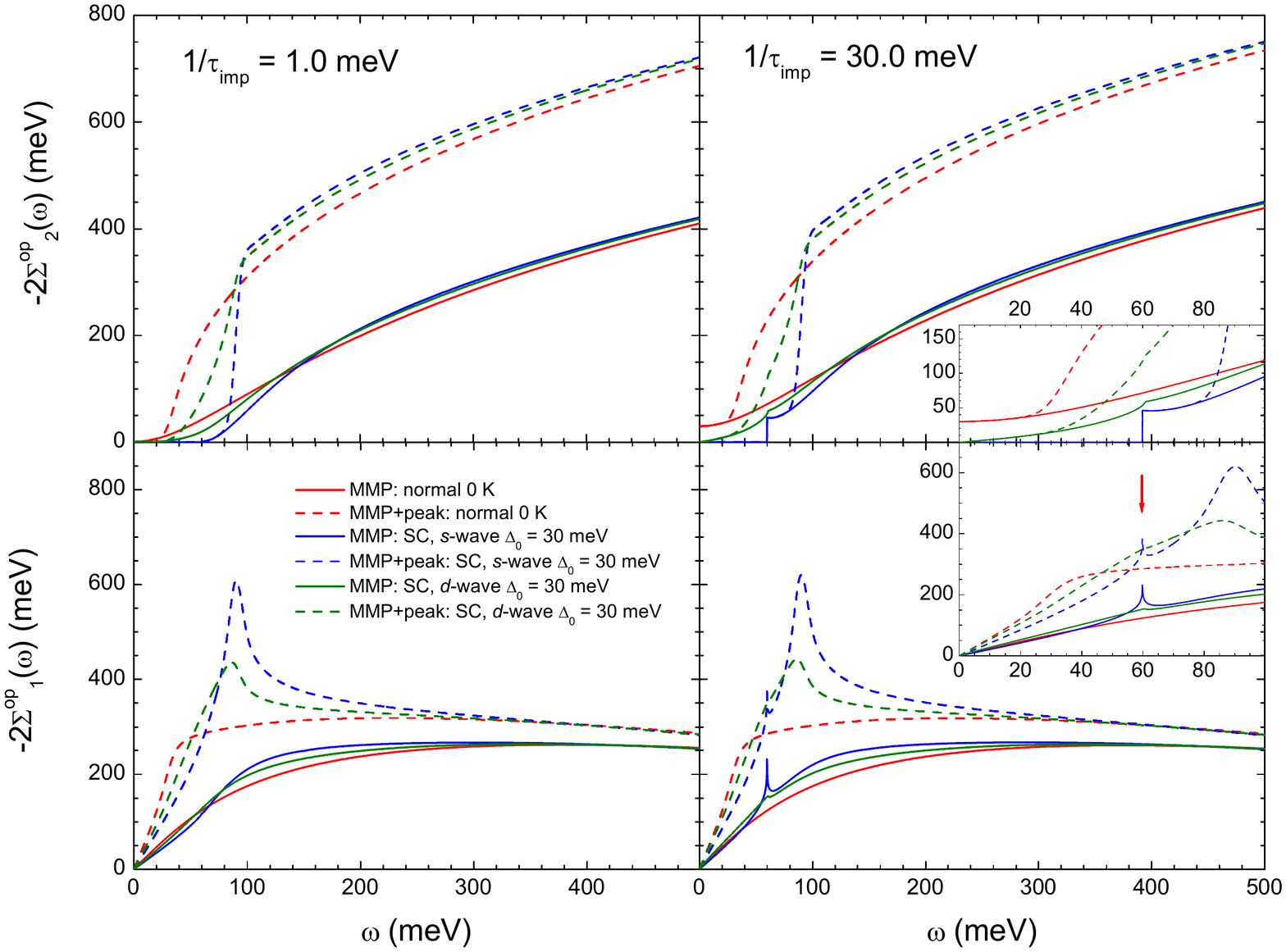}}%
  \vspace*{-0.7 cm}%
\caption{(Color online) The real and imaginary parts of the optical self-energy with two representative impurity scattering rates ($1/\tau_{imp}$), 1.0 meV (a clean limit) and 30.0 meV (a dirty limit). In the insets magnified views of the optical self-energy for the dirty limit case in low energy region are displayed. The red arrow indicates a sharp peak in the real part of the optical self-energy which is induced by the impurities.}
 \label{fig3}
\end{figure}

Now we obtained the real part from the imaginary of the optical self-energy using a Kramers-Kronig relation. In Fig. \ref{fig2} we display the obtained real parts of the optical self-energy [{\it i.e.} $-2\Sigma_1^{op}(\omega)$] in the lower left frame for the three cases with two input model $I^2\chi(\omega)$ along with their corresponding imaginary parts [{\it i.e.} $-2\Sigma_2^{op}(\omega)$] in the upper left frame. The results of normal state agree with the reported ones which were obtained in a similar situation\cite{hwang:2008a}. For two SC states with the MMP+peak $I^2\chi(\omega)$ there are prominent peaks near $\omega_p + 2\Delta_0$ ({\it i.e.} 90 meV) where the sharp rises are in the optical scattering rates. These peaks are originated from the sharp Gaussian peak in the input $I^2\chi(\omega)$. In the corresponding real part of the normal state we observe a kink-like feature (a abrupt slope change) near $\omega_p$ ({\it i.e.} 30 meV); the curve would have a logarithmic singularity in the slope at $\omega_p$ if we were using a pure delta function instead of the sharp Gaussian function\cite{carbotte:2005} with a finite width. The characteristic features of $I^2\chi(\omega)$ are clearly be seen in the optical self-energy.

Furthermore in the extended Drude formalism [see Eq. (\ref{eq4})] we obtained the optical conductivity with the plasma frequency ($\Omega_p$), 2.0 eV (or $\cong$ 16,130 cm$^{-1}$) and displayed the optical conductivity spectra for the three cases in the right frames [$\sigma_1(\omega)$ in the upper and $\sigma_2(\omega)$ in the lower] of Fig. \ref{fig2}. For this impurity-free ({\it i.e.} $1/\tau_{imp} = 0$) case we can see only the Holstein boson-assisted absorption piece in the real part of the optical conductivity, which is a fraction [$\lambda/(1+\lambda)$] of the total spectral weight, where $\lambda$ is the coupling constant (see Table \ref{table1}). The total spectral weight ($W_{s,total}$) is $\pi$/120 times the plasma frequency squared, {\it i.e.} $W_{s,total} \equiv \int_{0^{+}}^{\infty}\sigma_1(\omega)\: d\omega = (\pi/120)\: \Omega_p^2$, where $\sigma_1(\omega)$ is in units of $\Omega^{-1}$cm$^{-1}$ and $\Omega_p$ in units of cm$^{-1}$. The total spectral weights for all three cases should be the same since we used the same plasma frequency for the three cases in the extended Drude model. We compared the incoherent Holstein boson-assisted absorption spectral weights ($\cong 4.6 \times 10^6$ and $5.6 \times 10^6$ $\Omega^{-1}$cm$^{-2}$ obtained from the integration of the optical conductivity for the MMP and the MMP+peak, respectively) with the fractions [$\lambda/(1+\lambda) =$ 0.708 and 0.854 for the MMP and the MMP+peak, respectively] of the total spectral weight (6.7$\times 10^6$ $\Omega^{-1}$cm$^{-2}$), $\cong 4.7 \times 10^6$ and $5.7 \times 10^6$ $\Omega^{-1}$cm$^{-2}$ for the MMP and the MMP+peak, respectively. Here we took the coupling constants ($\lambda$) for the MMP and MMP+peak from Table \ref{table1}. The resulting incoherent spectral weights obtained from two different ways show good agreements for the both input $I^2\chi(\omega)$ cases. In $\sigma_2(\omega)$ we can see that the sharp Gaussian peak in the MMP+peak model appears as very prominent kinks (abrupt slope changes) near $\omega_p+2\Delta_0$ ({\it i.e.} 90 meV) for two SC states and a kink near $\omega_p$ ({\it i.e.} 30 meV) for the normal state. At the same frequencies where the kinks are we can see sharp rises in the real part of the optical conductivity, $\sigma_1(\omega)$. So we can clearly observe the characteristic features of $I^2\chi(\omega)$ in the optical conductivity.

\subsection{Impurity effects}

\begin{figure}[t]
  \vspace*{-0.3 cm}%
  \centerline{\includegraphics[width=6.0 in]{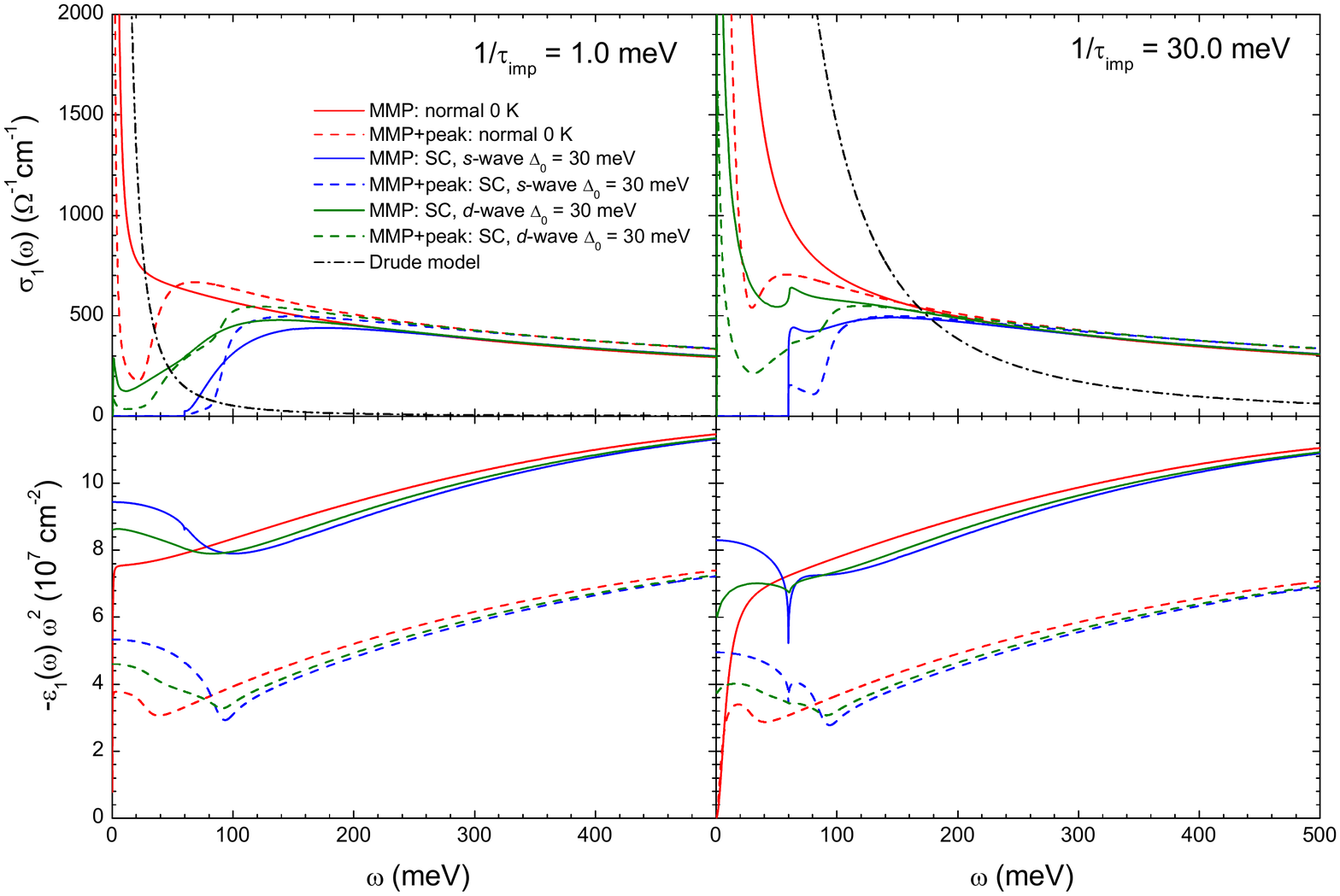}}%
  \vspace*{-0.7 cm}%
\caption{(Color online) Calculated real parts of the optical conductivity [$\sigma_1(\omega)$] and real parts of the dynamic dielectric function [$\epsilon_1(\omega)$] times -$\omega^2$ for all three cases with two different input $I^2\chi(\omega)$ with two representative impurity scattering rates ($1/\tau_{imp}$), 1.0 meV and 30.0 meV. We also show the real part of the optical conductivity spectra of the Drude modes (black dash-dotted lines) with corresponding impurity scattering rates.}
 \label{fig4}
\end{figure}

Now we added various impurity scattering rates (1, 5, 15, and 30 meV) to the impurity-free optical scattering rates as in Eq. (\ref{eq1}), (\ref{eq2}) and (\ref{eq3}) and used the same Kramers-Kronig relation to obtain the real parts of the optical self-energy. In Fig. \ref{fig3} we display the optical self-energy for two representative impurity scattering rates, $1/\tau_{imp} =$ 1.0 and 30.0 meV, which can be the clean ($1/\tau_{imp} \ll 2\Delta_0$) and the dirty ($1/\tau_{imp} \leq 2\Delta_0$) limit for $s$-wave and approximately $d$-wave superconductors, respectively. We observe clearly a very sharp peak (marked with a red arrow in the inset of the right lower frame) in the real part of the optical self-energy for the dirty limit case ($1/\tau_{imp} =$ 30.0 meV) at 2$\Delta_0$ ({\it i.e.} 60 meV) for $s$-wave SC case. At the same energy there is a sharp rise in the imaginary part of the optical self-energy as shown in the inset of the right upper frame. The sharp features seem to exist in the $d$-wave SC case but they are significantly suppressed; we can see it if we look at very carefully near 2$\Delta_0$ ({\it i.e.} 60 meV). This sharp feature is getting more pronounced as the impurity level increases, {\it i.e.} this feature can be induced by impurities. The sharp feature may not be observed in $d$-wave superconductors by infrared spectroscopic technique, which measures the averaged response over the anisotropic Fermi surface (or the $k$ space). But a similar impurity-induced sharp feature was observed in Bi$_2$Sr$_2$CaCu$_2$O$_{8+\delta}$ (a $d$-wave superconductor) by high-resolution ARPES\cite{zhang:2008,plumb:2010,kondo:2013}. This feature showed strong temperature and momentum dependencies. It also was interpreted theoretically as a result due to forward scattering in the $d$-wave cuprates\cite{hong:2014}.

\begin{figure}[t]
  \vspace*{-0.3 cm}%
  \centerline{\includegraphics[width=6.0 in]{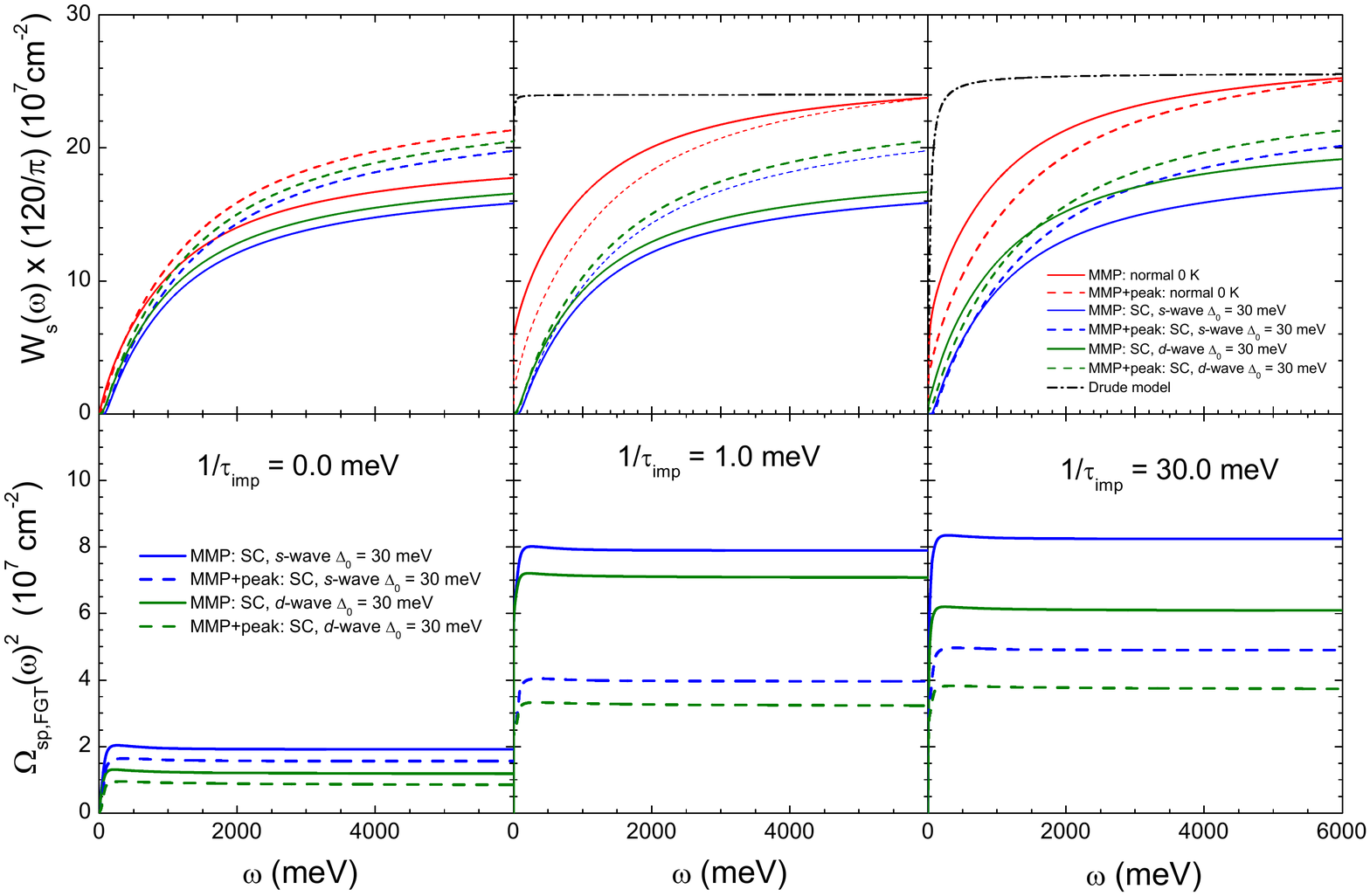}}%
  \vspace*{-0.7 cm}%
\caption{(Color online) The partial sum [$W_s(\omega)$] and the difference between normal and superconducting states, $W_{s,N}(\omega) - W_{s,SC}(\omega)$, for all three cases with two different input $I^2\chi(\omega)$ with three representative impurity scattering rates ($1/\tau_{imp}$), 0.0, 1.0 and 30.0 meV. We also show the partial sums of the Drude modes (black dash-dotted lines) with corresponding impurity scattering rate. We note that for $1/\tau_{imp} =$ 0 meV there is no spectral weight at finite frequency region.}
 \label{fig5}
\end{figure}

We also calculate the optical conductivity from the optical self-energy using the extended Drude formula [see Eq. (\ref{eq4})] with the same plasma frequency ($\Omega_p =$ 2.0 eV) at various impurity scattering rates (1, 5, 15, and 30 meV). In Fig. \ref{fig4} we display the calculated real parts of the optical conductivity and the real parts of the dynamic dielectric function for two representative impurity scattering rates, $1/\tau_{imp} =$ 1.0 and 30.0 meV. We also display the real parts (dot-dashed black lines) of the corresponding Drude optical conductivity in the upper frames. The Drude conductivity is completely coherent; there is no incoherent Holstein boson-assisted absorption since the charge carriers are essential free (or $\lambda = 0$). In other words the optical self-energy is zero, {\it i.e.} $\tilde{\Sigma}^{op}(\omega) = 0$, only elastic impurity scattering exists. For the normal cases (red solid and red dashed lines) we can see that the additional spectral weights appear at low frequency region in the optical conductivity compared with corresponding spectra for $1/\tau_{imp} = 0$ (see in the upper right frame of Fig. \ref{fig2}). Now the coherent portion of the total spectral weight [{\it i.e.} $W_{s,total}/(1+\lambda)$] also shows up in the conductivity spectra in a finite frequency range; the coherent spectral weight was confined at zero frequency for $1/\tau_{imp} = 0$. For $1/\tau_{imp} = 30.0$ meV (or the dirty limit) case we can see the gap features ({\it i.e.} sharp rises) clearly at $2\Delta_0$ ({\it i.e.} 60 meV) in $\sigma_1(\omega)$ of the $s$-wave superconductors. For the $d$-wave SC cases we also can see the gap features ({\it i.e.} sharp rises) at $2\Delta_0$ but the features are suppressed significantly by the anisotropic SC gap. For $1/\tau_{imp} = 1.0$ meV (or the clean limit) case those SC gap features cannot be seen very clearly. In the optical conductivity calculated by the reverse process we can observe the Gaussian peak in $I^2\chi(\omega)$ which appears as a dip near $\omega_p + 2\Delta_0$ ({\it i.e.} 90 meV) for both SC cases and near $\omega_p$ ({\it i.e.} 30 meV) for the normal case.

\begin{figure}[t]
  \vspace*{-0.3 cm}%
  \centerline{\includegraphics[width=4.5 in]{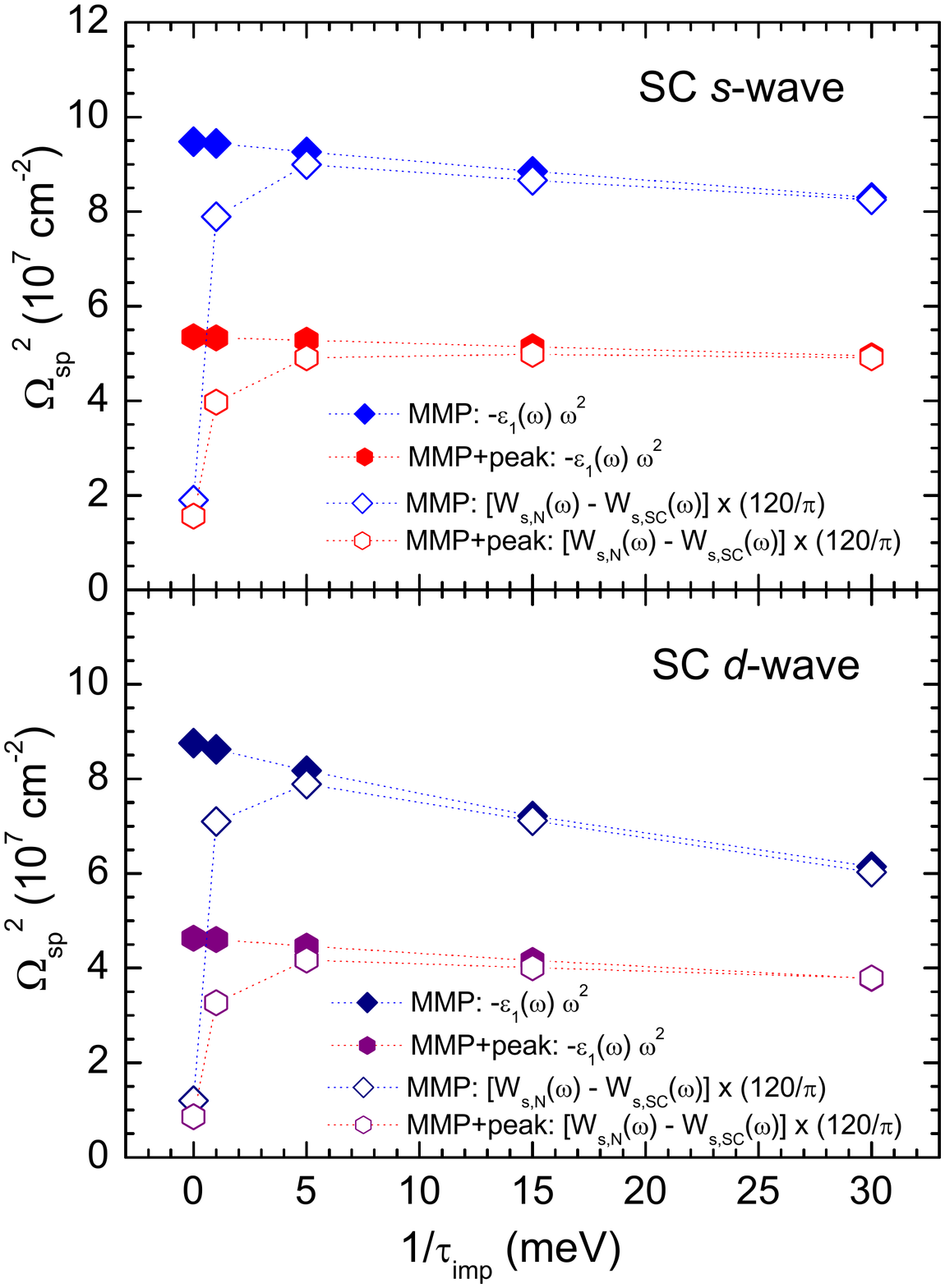}}%
  \vspace*{-0.7 cm}%
\caption{(Color online) Impurity dependent superfluid plasma frequencies obtained by two different methods (see in the text) for two different superconducting gap types ($s$-wave: upper frame and $d$-wave: lower frame) with the MMP and MMP+peak input $I^2\chi(\omega)$.}
 \label{fig6}
\end{figure}

In the lower frame of Fig \ref{fig4} we display $-\epsilon_1(\omega)\omega^2$ for all three cases. In principle this quantity at zero frequency gives the superfluid plasma frequency squared ($\Omega_{sp}^2$), {\it i.e.} $\lim_{\omega \rightarrow 0} [-\epsilon_1(\omega)\omega^2] = \Omega_{sp}^2$. We clearly can see the superfluid plasma frequency decreases as the impurity scattering rate increases for both superconducting cases; this result is expected for a superconductor moving into the dirty limit\cite{liang:1994,sun:1995}. The resulting superfluid plasma frequencies for various levels of impurities are calculated and shown in Fig. \ref{fig6}. For the normal case this quantity should be zero {\it i.e.} $\lim_{\omega \rightarrow 0} [-\epsilon_1(\omega)\omega^2] = 0$ since there is no superfluid current. In $-\epsilon_1(\omega)\omega^2$ the SC gap feature appears as a very sharp dip at $2\Delta_0$ ({\it i.e.} 60 meV). We also can clearly see the sharp Gaussian peak feature in the electron-boson spectral function near $\omega_p + 2\Delta_0$ ({\it i.e.} 90 meV) for both superconducting states and near $\omega_p$ ({\it i.e.} 30 meV) for the normal state, which appear as broad valleys.

\begin{figure}[t]
  \vspace*{-0.3 cm}%
  \centerline{\includegraphics[width=4.5 in]{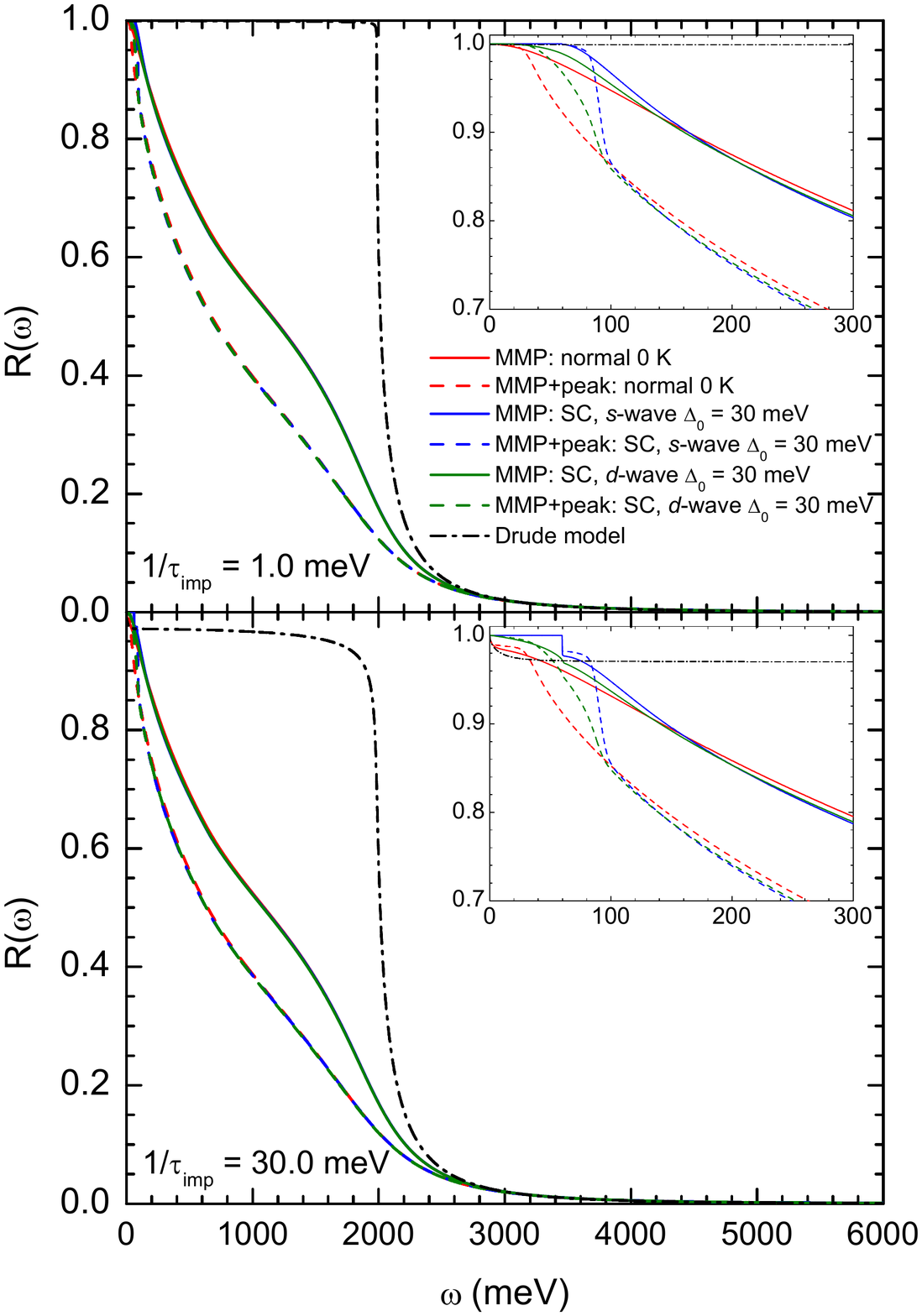}}%
  \vspace*{-0.7 cm}%
\caption{(Color online) Calculated reflectance spectra for all three cases with two different input $I^2\chi(\omega)$ with two representative impurity scattering rates ($1/\tau_{imp}$), 1.0 and 30.0 meV. We also show reflectance spectra of the Drude modes (black dash-dotted lines) with corresponding impurity scattering rates. In the insets we show magnified views of the reflectance in low frequency region.}
 \label{fig7}
\end{figure}

\begin{figure}
  \vspace*{-0.3 cm}%
  \centerline{\includegraphics[width=5.5 in]{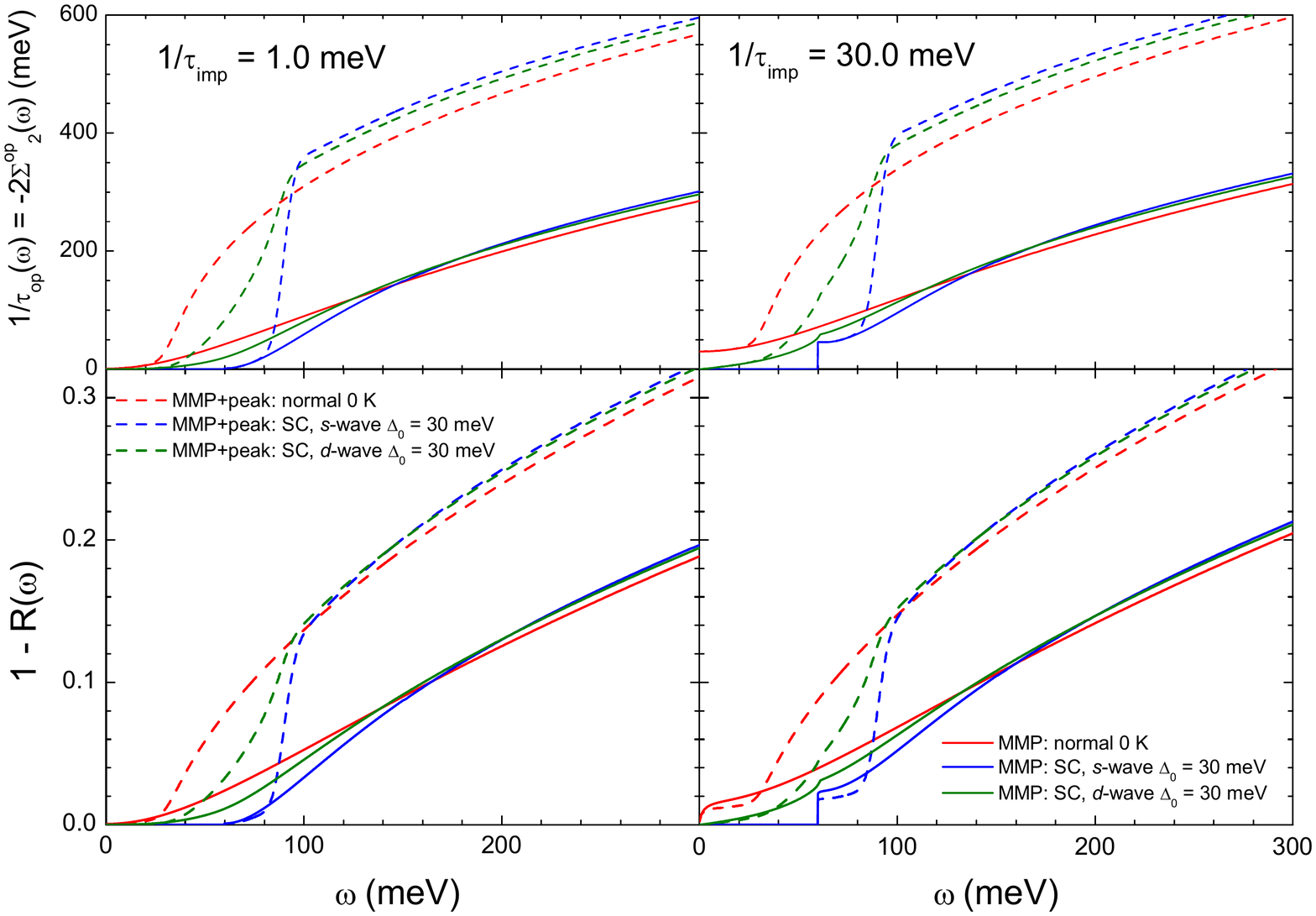}}%
  \vspace*{-0.7 cm}%
\caption{(Color online) Comparison of the optical scattering rate ($1/\tau^{op}(\omega)$) and $1 - R(\omega)$.}
 \label{fig8}
\end{figure}

Now we extract the superfluid plasma frequency with another method. In this method we monitor the spectral weight redistribution (or the missing spectral weight) when the system experiences a phase transition from normal to superconducting states. This method was originated from the Ferrel-Glover-Tinkham (FGT) sum rule\cite{glover:1956,ferrell:1958}. We denote the superfluid plasma frequency extracted using this method as $\Omega_{sp,FGT}$ and call as the FGT superfluid plasma frequency. We can describe $\Omega_{sp,FGT}$ in the FGT sum rule formalism as follows:
\begin{equation}\label{eq10}
\Omega_{sp,FGT}(\omega)^2 \equiv \frac{120}{\pi}[W_{s,N}(\omega) - W_{s,SC}(\omega)]
\end{equation}
where $W_{s,N}(\omega)$ and $W_{s,SC}(\omega)$ are the spectral weights of normal and superconducting states, respectively. We note that $\Omega_{sp,FGT}(\omega)$ is in units of cm$^{-1}$ and $W_s(\omega)$ in units of $\Omega^{-1}$cm$^{-2}$. For both normal and superconducting states the spectral weight is defined as $W_s(\omega) \equiv \int_{0^+}^{\omega}\sigma_1(\omega')\: d\omega'$. In the upper frames of Fig. \ref{fig5} we display the spectral weights ($W_s(\omega)$) for one normal and the two SC cases with two input model $I^2\chi(\omega)$ and three representative impurity scattering rates (0.0, 1.0, and 30.0 meV) in an extended spectral range (up to 6.0 eV). The corresponding $\Omega_{sp,FGT}^2(\omega)$ for the three representative impurity scattering rates are calculated and displayed in the lower frames. Interestingly, we have a finite superfluid plasmas frequency for $1/\tau_{imp} =$ 0.0 meV case, which indicates that some portion of the incoherent Holstein boson-assisted absorption is condensed into the superfluid state. At the same time we also can see that the total coherent portion of the spectral weight is smaller than the superfluid spectral weight for $1/\tau_{imp} =$ 0.0 meV case, {\it i.e.} $W_{s,total}\:[1/(1+\lambda)] < W_{s,FGT} \equiv (\pi/120)\: \Omega^2_{sp,FGT}$ for both input $I^2\chi(\omega)$. This supports the idea that some incoherent Holstein spectral weight was condensed into the superfluid density. Since there is an overshot in the FGT superfluid plasma frequency which extends up to around 2000 cm$^{-1}$ a reasonable estimate for the cutoff frequency for the FGT plasma frequency can be around 2000 cm$^{-1}$.

In Fig. \ref{fig6} we display the superfluid plasma frequencies obtained with two different methods as a function of the impurity scattering rate. The plasma frequencies obtained using $\Omega_{sp}^2 = \lim_{\omega\rightarrow 0}[-\epsilon_{1}(\omega)\:\omega^2]$ are shown with solid symbols and those obtained using $\Omega^2_{sp,FGT}(\omega) = \frac{120}{\pi}[W_{s,N}(\omega) - W_{s,SC}(\omega)]$ are displayed with open symbols. In low impurity scattering rates (or in the clean limit) below $1/\tau_{imp} =$ 5.0 meV $\Omega^2_{sp,FGT}(\omega)$ decreases very quickly with reducing the impurity scattering rate because the coherent portion is too narrow to be taken into account completely. This means that to get a good agreement between the two methods the system should contain enough level of impurities in order to make the coherent portion broad enough. But in high impurity scattering rates (or in the dirty limit) above 15.0 meV results from both methods agree each other quite well. We observe that the superfluid plasma frequency depends on the impurity scattering rate; as the impurity scattering rate increases the superfluid plasma frequency gradually decreases. The difference between the two superfluid plasma frequencies obtained by the two methods can be considered as an uncertainty for the FGT superfluid plasma frequency. We also observe that the superfluid plasma frequency of a system with the MMP+peak in $I^2\chi(\omega)$ shows more robust against the impurities than that with the MMP alone in $I^2\chi(\omega)$. Interestingly, the superfluid plasma frequencies of $s$-wave superconductors seem to be more robust against the impurities than $d$-wave ones.

Eventually, we calculated reflectance spectra [$R(\omega)$] using the Fresnel equation with the dynamic dielectric functions [see Eq. (\ref{eq8})] with $\epsilon_H = 1.0$. We note that the calculated reflectance spectrum is at normal incidence. In Fig. \ref{fig7} we display the calculated reflectance spectra for two representative impurity scattering rates ($1/\tau_{imp} =$ 1.0 and 30.0 meV). We also display Drude reflectance spectra (dash-dotted lines) with corresponding impurity scattering rates. We can see that correlated electron systems show much suppressed reflectance below the plasma frequency ($\Omega_p$ = 2.0 eV) compared with the free electron (or simple Drude) system. In the insets we show magnified views of reflectance spectra in low frequency region. For $1/\tau_{imp} =$ 30.0 meV the reflectance spectra of the $s$-wave SC case show clearly the superconducting gap feature at $2\Delta_0$ ({\it i.e.} 60 meV); below the energy the reflectance becomes essentially 1.0. For the $d$-wave SC case the gap feature is significantly suppressed; the reflectance becomes 1.0 at $\omega =$ 0. We also can see characteristic features which are caused by the input $I^2\chi(\omega)$. To see those features more clearly we compare reflectance spectra with the corresponding optical scattering rates; we display $1-R(\omega)$ and $1/\tau^{op}(\omega)$ in Fig. \ref{fig8}. Those two spectra look quite similar to each other. To show the similarity explicitly is not trivial since the reflectance is a quite complicated quantity theoretically. However the similarity can be understood roughly, as follows: in general at low frequency region a low scattering rate gives high conductivity which results in high $R(\omega)$ (or low $1-R(\omega)$) and vice versa for a high scattering rate. So roughly one can say that $1/\tau^{op}(\omega) \propto 1-R(\omega)$. Therefore one may be able to see characteristic features of $I^2\chi(\omega)$ as well as the gap features in the raw measured reflectance spectrum as one can see them in the optical scattering rate. We note that thermal excitations may smear the distinction of the features and/or shift the position of the SC gap features.

\subsection{Comparison with experimental data}

\begin{figure}
  \vspace*{-0.3 cm}%
  \centerline{\includegraphics[width=6.5 in]{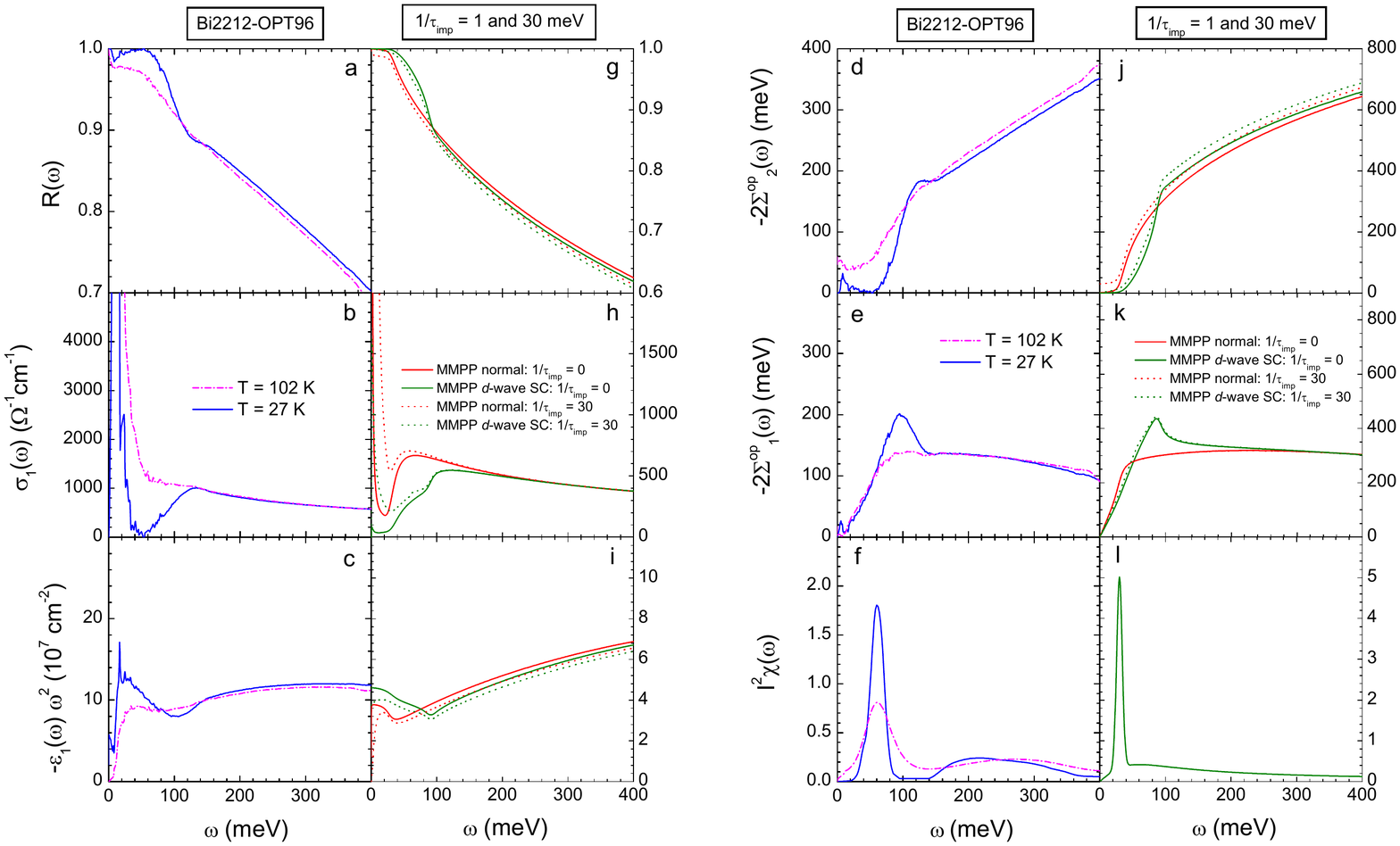}}%
  \vspace*{-1.5 cm}%
\caption{(Color online) Comparison of (a-f) measured reflectance and other optical constants obtained using the usual analysis of an optimally doped Bi$_2$Sr$_2$CaCu$_2$O$_{8+\delta}$ (Bi-2212) and (g-l) results obtained from the reverse process for $d$-wave SC case with the MMP+peak model $I^2\chi(\omega)$ for $1/\tau_{imp} =$ 1.0 (solid lines) and 30.0 meV (dotted lines).}
 \label{fig9}
\end{figure}

Now we compared measured experimental data with the corresponding optical data obtained by the reverse process. In Fig. \ref{fig9} we display measured reflectance data of a $d$-wave superconductor, optimally doped Bi$_2$Sr$_2$CaCu$_2$O$_{8+\delta}$ (Bi-2212) with $T_c =$ 96 K, at $T =$ 27 K (superconducting state) and 102 K (normal state) and optical constants obtained using the usual optical data analysis\cite{hwang:2007a} in (a-f) frames and the corresponding optical constants including reflectance obtained from the reverse process for $d$-wave SC case with the MMP+peak model $I^2\chi(\omega)$ for $1/\tau_{imp} =$ 1.0 and 30.0 meV in (g-l) frames. In (b) and (h) frames we can see that the results with $1/\tau_{imp} = 30.0$ meV show better agreements to the experimental date, which indicates that certain impurities caused by defects and/or thermal excitations exist in the experimental data. The experimental data would show temperature smearing effects from thermal excitations. Overall all optical data displayed show quite good agreements even though experimental data are at finite temperatures while our calculations are done at zero temperature. We note that the electron-boson spectral function of Bi-2212 shows an additional intrinsic temperature dependent evolution as well\cite{carbotte:2011,hwang:2007}. We also note that in the calculations we used the same input $I^2\chi(\omega)$ for both normal and superconducting states and for $1/\tau_{imp} =$ 1.0 and 30.0 meV. We note that the real part of the optical self-energy shows relatively small impurity dependence. Actually for the normal case $-2\Sigma_1^{op}(\omega)$ should not depend on the impurity level at all since a frequency independent impurity scattering rate is added and its Kramers-Kronig counterpart is essentially zero. However for superconducting cases since the impurity scattering rates show frequency dependence near $2\Delta_0$ their Kramers-Kronig counterparts are not zero any more. This gives the difference between $-2\Sigma_1^{op}(\omega)$ obtained with different impurity scattering rates near $2\Delta_0$. While the positions of the peaks of calculated and measured $-2\Sigma_1^{op}(\omega)$ have similar energies ({\it i.e.} 100 meV in frame (e) and 90 meV in frame (k)) the peaks of two (experimentally obtained and the input) $I^2\chi(\omega)$ are located at quite different energies ({\it i.e.} 60 meV in frame (f) and 30 meV in frame (l)). This difference comes from different superconducting energy gaps ({\it i.e.} $\Delta_0 \simeq$ 20 meV for experimental data and $\Delta_0 =$ 30 meV for the calculated ones). As we mentioned in section 3.2 the peak in $-2\Sigma_1^{op}(\omega)$ is located at $\omega_p+2\Delta_0$. So we can understand that 100 meV $\simeq$ 60 meV + 2$\times$20 meV for the peak in experimental data (frame (e)) and 90 meV = 30 meV + 2$\times$30 meV for the peak in calculated one (frame (k)). These good agreements confirm that our reverse process is quite reliable.

\section{Conclusions}

In this study we have performed a reverse process of the usual process for optical data analysis. The reverse process is as follows: We started with the input model electron-boson spectral function, $I^2\chi(\omega)$. We obtained the imaginary part of the optical self-energy using Allen's formulas, the real part of the optical self-energy using a Kramers-Kronig relation, the optical conductivity using the extended Drude model, other optical constants using well-known relations between optical constants, and eventually reflectance using the Fresnel equations. We applied this reverse process to three cases (one normal and $s$- and $d$- superconducting cases with the superconducting gap $\Delta_0 =$ 30 meV) with two (MMP and MMP+peak) input model $I^2\chi(\omega)$ which are based on the reported experimental results. Since impurities are not avoidable experimentally we included various levels of impurities (from the clean to the dirty limit: $1/\tau_{imp} =$ 0 $\sim$ 30.0 meV) in the optical scattering rate and performed the same reverse process to investigate impurity dependent optical properties. In the clean limit (for example, $1/\tau_{imp} =$ 1.0 meV) for the both superconducting cases we do not observe the superconducting gaps clearly but still the characteristic feature ({\it i.e.} the sharp Gaussian peak and the MMP model) of $I^2\chi(\omega)$ appears definitely in all optical constants including reflectance. For the normal case we also are able to see the feature of $I^2\chi(\omega)$ in the optical constants including reflectance. In the dirty limit (for example, $1/\tau_{imp} =$ 30.0 meV) the superconducting gap appears definitely in $s$-wave superconductors but the gap feature is suppressed significantly in the $d$-wave one. In $d$-wave superconductors the optical constants are averaged response signals over the anisotropic Fermi surface which causes the suppression of those features. Furthermore in the dirty limit the characteristic features (the Gaussian peak and the MMP model) of $I^2\chi(\omega)$ also can be seen clearly in the optical constants and measured reflectance spectrum. We also find that in the clean limit the superfluid density obtained by the FGT sum rule cannot capture whole coherent electrons which participate in the superconductivity because the coherent absorption peak is too narrow. To get accurate superfluid density the SC system should be in a dirty limit. Our study shows that the superfluid density obtained by the FGT sum rule becomes quite accurate when $1/\tau_{imp}$ is large enough, above 15 meV, which is comparable to the superconducting gap, $\Delta_0$ (30.0 meV). We observe that the superfluid density decreases as the impurity level increases for both SC cases. We also observe that the $s$-wave superconductor is more robust against the impurities than the $d$-wave one. Our results will help to understand the usual optical analysis process thoroughly as well. Since the optical spectra show clearly the mediated boson features of the boson-exchange superconductors collecting experimental data at various (temperature, doping, magnetic field and so on) conditions and obtaining $I^2\chi(\omega)$ applying the usual process can be a first step to reveal the microscopic origin of the exchange boson.

\ack
The author acknowledges financial support from the National Research Foundation of Korea (NRFK Grant No. 2013R1A2A2A01067629).

%
% bibliography
%

\section*{References}
\bibliographystyle{unsrt}
\bibliography{bib}% Produces the bibliography via BibTeX.

\end{document}